\documentclass[aps,prd,twocolumn,superscriptaddress,showpacs]{revtex4-1}

\usepackage{amsmath}
\usepackage{graphicx}
\usepackage{epstopdf}
\usepackage{color}
\usepackage{slashed}
\usepackage{CJK}
\definecolor{purple}{rgb}{0.5,0,0.5}
\definecolor{blue}{rgb}{0.0,0,0.9}
\definecolor{prdblue}{rgb}{0.133,0.118,0.498}

\usepackage[colorlinks=true, pdfstartview=FitV, linkcolor=prdblue, citecolor= prdblue, urlcolor=prdblue]{hyperref}

\begin{document}

\begin{CJK*}{GBK}{song}

\title{Thermodynamics of 2+1 Flavor Polyakov-Loop Quark-Meson Model under External Magnetic Field}

\author{Xiang Li}
\affiliation{Department of Physics and State Key Laboratory of
Nuclear Physics and Technology, Peking University, Beijing 100871,
China}
\affiliation{Collaborative Innovation Center of Quantum Matter, Beijing 100871, China}
\author{Wei-jie Fu}
\affiliation{School of Physics, Dalian University of Technology, Dalian 116024, China}
\author{Yu-xin Liu}
\email[Corresponding author: ]{yxliu@pku.edu.cn}
\affiliation{Department of Physics and State Key Laboratory of
Nuclear Physics and Technology, Peking University, Beijing 100871,
China}
\affiliation{Collaborative Innovation Center of Quantum Matter, Beijing 100871, China}
\affiliation{Center for High Energy Physics, Peking University, Beijing 100871, China}

\date{\today}

\begin{abstract}
We study the thermodynamics of QCD system under external magnetic field via the 2+1 flavor Polyakov-loop quark-meson model. To incorporate quantum and thermal fluctuations, the functional renormalization group approach is implemented in our work. Pressure, entropy density, magnetic susceptibility and other thermodynamic quantities are calculated and analyzed to investigate the effect of magnetic field on the QCD system.
The calculated results are in reasonable agreement with lattice QCD simulations and perturbation theory.
We then give an intuitive picture for the response of QCD system to the magnetic field.
\end{abstract}

%\pacs{12.38.Aw, 21.65.Qr, 12.39.Fe, 11.10.Wx}
%
%\pacs{
%12.38.Aw, %  General properties of QCD (dynamics, confinement, etc),
%25.75.Nq  %	Quark deconfinement, quark-gluon plasma production, and phase transitions
%%12.38.Mh, %  quark-gluon plasma,
%11.10.Wx, %	Finite-temperature field theory
%12.38.Lg,  %	Other nonperturbative calculations
%}

\maketitle

\section{Introduction}

Researches on the phase transitions of strong interaction matter (QCD phase transitions) have attracted great attentions in recent years (for example, see Refs.~\cite{Gyulassy:2005NPA,Aoki:2006NAT,Blaizot:2007Review,Wambach:2009Review,Gupta:2011Sci,Fukushima:201112Reviews,Rehberg:1996PRC,Stephanov:2007fk,McLerran:2007NPA,Skokov:2012PRD,Mizher:2010PRD,Mao:2010JP,Schaefer:2007PRD,Liu:Phenomodel,Qin:2011PRL,Fischer:201134,Pawlowski:2013,Philipsen:2013,Borsanyi:2010JHEP,Liu:2011PRD,Gavai:2011PLB,Philipsen:2014PRL,DElia:20101PRD,Bazavov:2012PRD,Ejiri:2013PRL,Bali:2012JHEP,Bali:2014JHEP,Bonati:2014PRD,Levkova:2014PRL,Bornyakov:2014PRD,Jiang:2013PRD,Ke:2014PRD,Endrodi:2013JHEP,Gao:20168PRD,Fischer:2018,Zhuang:2012}) since they may shed light on revealing the process and the nature of the early universe evolution.
It is known that not only during the early stage of the universe evolution but also in the modern heavy-ion collision experiments, there exists strong magnetic field~\cite{Vachaspati:1991PLB,Skokov:2009IJMPA},
which may influence  the property of the strong interaction matter drastically.
The effects of the magnetic field have naturally been investigated intensively.
As a consequence, it has arisen some interesting questions,
for instance, the chiral magnetic effect~\cite{Kharzeev:2008NPA,Fukushima:2008PRD,Buividovich:2009PRD,Fukushima:2010PRD,Fu:2010IJMPA,Hou:2011JHEP}  and the magnetic catalysis effect~\cite{Klimenko:1992Phys,Suganuma:1991AnP,Miransky:2002PRD,Smilga:1997PLB,Kabat:2002PRD,Chernodub:2011PRL,DElia:20101PRD}.

However, recent lattice QCD simulation results show us that, when considering the interaction more sophisticatedly, the magnetic field does not increase the critical temperature for the chiral symmetry to be restored~\cite{Bali:2012JHEP,Bali:2014JHEP,Bornyakov:2014PRD}, which is referred to as  the inverse magnetic catalysis (magnetic inhibition) effect.
Plenty of theoretical works have then been accomplished to clarify which of the two completely opposite effects is correct, or to what extent each of the effects works~\cite{Skokov:2012PRD,Mizher:2010PRD,Bali:2014JHEP,Bornyakov:2014PRD,Fayazbakhsh:2011PRD,Farias:2014PRC,Ayala:2012PRD,Fukushima:2013PRL,Kojo:2013PLB,Bruckmann:2013JHEP,Chao:2013PRD,Ilgenfritz:2014PRD,Fraga:2014PRD,Andersen:2014JHEP,Ruggieri:2014PLB,Fischer:2014DSE,Bali:2014JHEP,Orlovsky:2014PRD,Ferrer:2014PRD,Kamikado:2014JHEP,Kamikado:2015JHEP,Ayala:2014PRD,Ferreira:2014PRD1,Ferreira:2014PRD2,Huang:2014PRD,Braun:2016PLB,Fu:2017PRD,Ayala:2015PRD1,Sadooghi:2014PRD,Andersen:2015JHEP,Zong:2015PRD,Ferrer:2015PRD,Mamo:2015JHEP,Pawlowski:2015PRD,Loewe:2015IJMPA,Ayala:2015PRD2,Ferreira:2015,Liu:2015CTP,Aoki:2016PLB,DFHou:2015PRD}, even explore the underlying mechanism of the effects from a fundamental point of view (see, for example, Refs.~\cite{Klimenko:1992Phys,Miransky:2002PRD,Fukushima:2013PRL,Chao:2013PRD,Orlovsky:2014PRD,Ferrer:2014PRD,Kamikado:2014JHEP,Kamikado:2015JHEP,Ayala:2014PRD,Huang:2014PRD,Braun:2016PLB,Fu:2017PRD,Sadooghi:2014PRD,Ayala:2015PRD1,Ferrer:2015PRD,Pawlowski:2015PRD,Andersen:2015JHEP,Zong:2015PRD,Ayala:2015PRD2,Loewe:2015IJMPA} and recent reviews~\cite{Andersen:2014JHEP,Shovkovy:2015Review,Andersen:2016Review}).
Even though there is still no definitive conclusion, it has been recognized commonly that the external magnetic field produces a spatial dimension reduction for charged particles (see, e.g., Refs.~\cite{Miransky:2002PRD,Fukushima:2013PRL,Pawlowski:2015PRD,Zong:2015PRD,Ayala:2015PRD2,Liu:2015CTP,Shovkovy:2015Review}).
However one has not yet arrived at a common idea on the effect of the dimension reduction, for instance,  Refs.~\cite{Zong:2015PRD,Liu:2015CTP,Shovkovy:2015Review} claim that the dimension reduction strengthens quark condensate and leads to magnetic catalysis at low temperature but Refs.~\cite{Fukushima:2013PRL,Ayala:2015PRD2} say that the low dimensionality may favor the inverse magnetic catalysis.
It is then imperative to clarify the effect of the external magnetic field, especially with an intuitive or a pedagogic picture.

It has been well known that the equation of state (EoS) and the entropy density can represent well the properties of a matter. Then the EoS and the entropy density of the strong interaction matter in an external magnetic field have been studied with the lattice QCD simulations~\cite{Bali:2014JHEP,Bonati:2014PRD,Levkova:2014PRL}, Hadron Resonance Gas (HRG) model \cite{Endrodi:2013JHEP}, quark-meson (QM)
model~\cite{Kamikado:2014JHEP,Kamikado:2015JHEP,Mitter:2014PRD} and Polyakov-loop improved quark-meson (PQM)
model~\cite{Skokov:2012PRD,Mizher:2010PRD,Mao:2010JP,Schaefer:2007PRD,Tawfik:2014PRC,Andersen:2014JHEP,Herbst:2014PLB,Fabian:2017PRD,Fu:2018,Fu:20182}. However,
the mechanism of the magnetic field effect on the QCD system has not yet been fully understood by analyzing the thermodynamical properties concretely. For instance, the responses of the quark condensate, pressure and entropy density of QCD system to the magnetic field at different temperature regions and the underlying physics picture  remain unclear, which all require more investigations. We then investigate the effect of the magnetic field on the thermodynamic properties of the strong interaction matter and try to give an intuitive view point of the
mechanism of the magnetic catalysis effect in some simple models in this paper.

In this work, we take the (2+1) flavor Polyakov-loop quark-meson model~\cite{Skokov:2012PRD,Mizher:2010PRD,Mao:2010JP,Schaefer:2007PRD,Tawfik:2014PRC,Andersen:2014JHEP,Herbst:2014PLB,Fabian:2017PRD,Fu:2018,Fu:20182} to approximate the QCD system. Compared with previous works~\cite{Skokov:2012PRD,Mizher:2010PRD,Schaefer:2007PRD,Kamikado:2014JHEP,Kamikado:2015JHEP,Andersen:2014JHEP}, the effects of  Polyakov-loop and strange quarks are taken into considerations.
Adding Polyakov-loop can imitate the effect of confinement and including strange quark brings more mesons
(such as kaon) into the system so that the system is more close to the real one at the high energy scale.
Besides using a more realistic model, functional renormalization group (FRG) approach~\cite{Wetterich:1993Review,Litim:2000PLB,Jan:2007Review,Gies:2012Review} is implemented to investigate the thermodynamics of QCD system at finite temperature and magnetic field.
FRG approach can incorporate the fluctuations of quarks and mesons thus goes beyond the mean-field approximation.
It has then been shown powerful to solve non-perturbative problems.

The remainder of this paper is organized as follows. In Sec.~\ref{Sec:pqm}, we introduce briefly the (2+1) flavor PQM model. In Sec.~\ref{Sec:frg}, we depict the framework of the FRG approach.
In Sec.~\ref{Sec:res}, we describe our calculation procedures and the obtained results,
and provide an intuitive picture for the magnetic effect.
In Sec.~\ref{Sec:sum}, we give our summary and some remarks.

\section{2+1 FLAVOR PQM MODEL}\label{Sec:pqm}

The Polyakov-loop quark-meson model is an effective approximation of QCD
system~\cite{Skokov:2012PRD,Mizher:2010PRD,Mao:2010JP,Schaefer:2007PRD,Tawfik:2014PRC,Andersen:2014JHEP,Herbst:2014PLB,Fabian:2017PRD,Fu:2018,Fu:20182}.
The Lagrangian (in Euclidean space) of the (2+1) flavor PQM model under an external magnetic field and at vanishing chemical potential can be simply given as
\begin{eqnarray}\label{Eq:lag}
\mathcal{L}=\ &&\bar\psi(\slashed D-i\gamma_4A_4+g\Sigma_5)\psi \nonumber\\
&&+\ {\rm Tr}[\mathcal{D}_\mu\Sigma\cdot (\mathcal{D}_\mu\Sigma)^\dagger]+\tilde{U}(\Sigma)+V_{poly}(\Phi)\, .
\end{eqnarray}
The first line represents quark sector, $\psi$ is the quark field with 3 flavors (u, d, s) and 3 colors. $D_\mu=\partial_\mu-iqA_\mu^{\rm em}$ is the covariant derivative coupled to quarks
with $q={\rm diag}(\frac{2}{3}e,-\frac{1}{3}e,-\frac{1}{3}e)$ being a diagonal matrix encoding the charge for each quark. We assume that the external magnetic field $B$ is homogeneous and along the z-direction, the corresponding field $A_{\mu}^{\rm em}$ can be written as
\begin{eqnarray}
A_{\mu}^{\rm em}=(0,Bx,0,0)\, .
\end{eqnarray}
The Yukawa term $g\bar\psi\Sigma_5\psi$ is also included to represent the interaction between quarks and mesons.

The first two terms in the second line of Eq.~(\ref{Eq:lag}) represent the meson sector.
Meson fields can be combined to form matrices, which read
\begin{eqnarray}
\Sigma=&&\ \ \sum_{a=0}^8\ (\sigma_a+i\pi_a)T^a\, ,\nonumber\\
\Sigma_{5}=&&\ \ \sum_{a=0}^8\ (\sigma_a+i\gamma_5\pi_a)T^a\, ,
\end{eqnarray}
where $\sigma_a$ and $\pi_a$ are the scalar and pseudo-scalar meson nonets. $T^a$ is the ${\rm SU(3)}$ generator in flavor space, and $T^{0}= \sqrt{\frac{1}{6}} I_{3 \times 3}^{} $. The derivative term for mesons is a little complicated which is defined as
\begin{eqnarray}
\mathcal{D} _\mu\Sigma=\partial_\mu\Sigma-iA_\mu^{\rm em}[q,\Sigma]\, .
\end{eqnarray}
$\tilde{U}(\Sigma)$ is the meson potential which contains several parts and reads
\begin{eqnarray}
\tilde{U}(\Sigma)=U(\rho_1,\rho_2)-h_x\sigma_x-h_y\sigma_y-c_a\xi\, ,
\end{eqnarray}
where $U(\rho_{1}^{},\rho_{2}^{})$ is the symmetric term under ${\rm U_{V}(3)\times U_{A}(3)}$ transformation.
$\rho_{1}^{}$ and $\rho_{2}^{}$ are two invariants constructed from $\Sigma$ as
\begin{eqnarray}
\rho_{1}^{}=\ &&{\rm Tr}\big{[} \Sigma\cdot\Sigma^{\dagger} \big{]}\, ,\nonumber\\
\rho_{2}^{}=\ &&{\rm Tr}\big{[} (\Sigma\cdot\Sigma^{\dagger} - \frac{1}{3}\rho_{1}^{})^{2} \big{]} \, .
\end{eqnarray}
$h_{x} \sigma_{x}$ and $h_{y} \sigma_{y}$ are the explicit symmetry breaking terms and they are added to
reduce the ${\rm SU_{V}(3)}$ symmetry to ${\rm SU_{V}(2)}$ which is approximately conserved in QCD. $\sigma_{x}$ and $\sigma_{y}$ are related to the meson fields via a rotation
\begin{eqnarray}
\begin{pmatrix}
\sigma_{x}\\
\sigma_{y}
\end{pmatrix}
=\frac{1}{\sqrt{3}}
\begin{pmatrix}
\sqrt{2} & 1\\
1 & -\sqrt{2}
\end{pmatrix}
\begin{pmatrix}
\sigma_{0} \\
\sigma_{8}
\end{pmatrix}  .
\end{eqnarray}
$\xi$ is the  Kobayashi--Maskawa--'t Hooft determinant which is used to imitate the ${\rm U_{A}}(1)$ anomaly effect.
It breaks the ${\rm U_A(1)}$ symmetry explicitly and reads
\begin{eqnarray}
\xi={\rm det}(\Sigma)+{\rm det}(\Sigma^\dagger)\, .
\end{eqnarray}

In order to take color confinement effect into consideration, a background gluon field $A_{4}$ and its potential  $V_{poly}({\rm \Phi})$ are added into the Lagrangian. $\Phi$ is called Polyakov-loop and is defined as
\begin{eqnarray}
\Phi=\frac{1}{N_c}{\rm Tr\ P\ exp}(i\int_0^\beta d\tau A_4)\, ,
\end{eqnarray}
where P is path order operator. $V_{poly}({\rm \Phi})$ is a phenomenological potential obtained by fitting lattice QCD data and it's used to represent the gluon's contribution to the Lagrangian. The specific form of the $V_{poly}$ and the parameters therein will be discussed in Sec.~\ref{Sec:frgnum}.
%
%Furthermore, the chemical potential of each flavor can in principle be different from each other,
%However one usually takes $\mu_{u}^{} = \mu_{d}^{} = \mu_{s}^{} = \mu$ for simplicity.
%
\section{FUNCTIONAL RENORMALIZATION GROUP APPROACH}\label{Sec:frg}
\subsection{THEORETICAL FRAMEWORK}
After the Lagrangian is introduced, several approaches can be taken to calculate the effective action $\Gamma$ of the system. For example, mean-field approximation is a widely used method. It treats meson fields as background and neglects their fluctuations. However, it has been shown that meson's fluctuations are important for investigating QCD system at low temperature~\cite{Kamikado:2015JHEP}.
As a consequence, we take the functional renormalization group (FRG) approach instead in this article.
With the FRG approach, one can incorporate the fluctuations of mesons as well as quarks.

FRG approach is usually employed to solve non-perturbative problems (see, e.g.,  Refs.~\cite{Wetterich:1993Review,Litim:2000PLB,Jan:2007Review,Gies:2012Review} for details).
It is well known that the $\Gamma$ can be expanded in terms of loops and FRG approach implements a momentum-dependent mass term $R_{k}(p)$ for each propagator in the loops.
These modified propagators get highly suppressed at small momentum region $(p<k)$
while are left unchanged at large momentum region $(p>k)$.
When $k$ is near the cutoff $\Lambda$, all the loop integrals are zero and only bare action contributes to
the effective action, i.e., $\Gamma_{k=\Lambda}=S_{{\rm Bare}}$.
When $k$ approaches zero, all the loop integrals are restored and we get the full effective action as
$\Gamma_{{\rm Full}}= \Gamma_{k=0}$.
These ideas can be formulated into a functional differential equation, which reads
\begin{eqnarray}
\partial_{k} \Gamma_{k} =\frac{1}{2}{\rm Tr}\Bigg[\frac{\partial_{k} R_{k}^{B}}{\Gamma^{2B}_{k} + R_{k}^{B}}\Bigg]
-{\rm Tr}\Bigg[\frac{\partial_{k} R_{k}^{F}}{\Gamma^{2F}_{k} + R_{k}^{F}}\Bigg]\, ,
\label{Eq:wet}
\end{eqnarray}
where $\Gamma_k^{2B}$ ($\Gamma_k^{2F}$) denotes the functional derivative of $\Gamma_{k}$ with respect to boson (fermion) field. $R_{k}^{B,F}$ is the momentum-dependent mass term assigned to each particle.
${\rm Tr}$ represents trace operation in all spaces (momentum, flavor, color, $\cdots$).

One of the main characters of FRG approach is its differential structure. $S_{\rm Bare}$ serves as the initial condition for Eq.~(\ref{Eq:wet}). We can study the whole system by solving the differential equations instead of doing loop integrals or solving integral equations.

However, it is extremely hard to solve the functional equations exactly.
Some simplifications are necessary to solve the problems practically.
We will briefly review the necessary truncations and the derivations for the PQM model as follows.
The details can be found in Refs.~\cite{Kamikado:2015JHEP,Mitter:2014PRD,Andersen:2014JHEP,Herbst:2014PLB,Fabian:2017PRD,Fu:2018,Fu:20182}.

We employ the so called local potential approximation (LPA). In this approximation, all the non-local terms and anomalous dimensions of the fields are omitted.  The truncated $\Gamma_k$ now reads
\begin{eqnarray}\label{Eq:effgamma}
\Gamma_{k} =&& \; \int d^{4}x\ \bar\psi(\slashed D - i \gamma_{4}^{} A_{4} + g\Sigma_{5})\psi \nonumber\\
&& \;\; +\,{\rm Tr}[\mathcal{D}_\mu\Sigma\cdot (\mathcal{D}_{\mu} \Sigma)^{\dagger}] + U_{k}(\rho_{1}^{} , \rho_{2}^{} ) \nonumber\\ && \;\; -\,h_{x} \sigma_{x} - h_{y} \sigma_{y} - c_{a} \xi + V_{poly}(\Phi) \, ,
\end{eqnarray}
where only $U_k(\rho_1,\rho_2)$  flows with scale $k$ and other terms are the same as Eq.~(\ref{Eq:lag}).
 Substituting Eq.~(\ref{Eq:effgamma}) into Eq.~(\ref{Eq:wet}) and implementing the optimized regulators taken in Refs.~\cite{Skokov:2012PRD,Andersen:2014JHEP,Kamikado:2015JHEP,Litim:2000PLB}, one can obtain an analytic flow equation for the $U_{k}$ as
\begin{eqnarray}\label{Eq:effpotflow}
\partial_kU_k=&&\ \frac{k^4}{12\pi^2}\Big\{\sum_b\alpha_b(k)\frac{1}{E_b}[1+2n_b(E_b)]\nonumber \\ &&-\sum_{f=u,d,s}\alpha_f(k)\frac{1}{E_f}[1-2\tilde{n}_f(E_f,\Phi)]\Big\}\, .  \qquad
\end{eqnarray}
For the definitions of $\alpha_{b(f)},E_{b(f)}$, see Ref.~\cite{Kamikado:2015JHEP}. $n_b$ is the ordinary boson distribution function while $\tilde{n}_{f}$ is  the Polyakov-loop modified fermion distribution function which reads
\begin{eqnarray}\label{Eq:polyfermion}
\tilde{n}_f(E_f,\Phi)=\frac{1+2\Phi e^{\beta E_f}+\Phi e^{2\beta E_f}}{1+3\Phi e^{\beta E_f}+3\Phi e^{2\beta E_f}+e^{3\beta E_f}}\, .  \qquad
\end{eqnarray}
At low temperature, $\Phi\sim 0$ and Eq.~(\ref{Eq:polyfermion}) is reduced to
\begin{eqnarray}
\tilde{n}_{f}^{} =\frac{1}{e^{3\beta E_{f}}+1}\, .
\end{eqnarray}
This distribution implies that the average energy of a quark is nearly one third of an ordinary free fermion. In other words, quarks are statistically confined at low temperature. At high temperature, $\Phi\sim1$ and $\tilde{n}_f$ is reduced to the ordinary fermion distribution function.

There are usually two algorithms to solve Eq.~(\ref{Eq:effpotflow}). One is the so called grid method.
It discretizes the arguments of $U_{k}$ (i.e., the $\rho_{1}^{},\rho_{2}^{}$ and $\Phi$ mentioned above) on a multi-dimensional grid, then Eq.~(\ref{Eq:effpotflow}) can be transformed into a set of coupled ordinary differential equations.
However, grid method requires usually much numerical effort, especially when the number of the arguments increases.
So Taylor expansion method can be employed instead.
Taylor expansion method expands the $U_{k}(\rho_1,\rho_2)$  around the minimums $\rho_1(k),\rho_2(k)$ and
the series is truncated to a certain order. The expansion series reads
\begin{eqnarray}\label{Eq:taylor}
U_{k} = \sum_{i+2j=0}^n\frac{a_{i,j}(k)}{i!j!}[\rho_1-\rho_{1}(k)]^i[\rho_2-\rho_{2}(k)]^j\, .
\end{eqnarray}
Setting $n=5$ has been shown to be sufficient for numerical convergence~\cite{Fu:20182}. A set of  coupled ordinary differential equations for $a_{i,j}$ and $\rho_{i}^{}(k)$ can then be obtained by substituting Eq.~(\ref{Eq:taylor}) into Eq.~(\ref{Eq:effpotflow}). The flow equations read simply (specific form can be seen in Refs.~\cite{Andersen:2014JHEP,Kamikado:2015JHEP})
\begin{eqnarray}\label{Eq:stru}
k\frac{da_{i,j}}{dk}=\beta_{i,j}(a_{m,n},k,T,eB)\, ,
\end{eqnarray}
and there are many numerical methods to solve these equations.

In short, effective action $\Gamma_k$ can be parameterized into a set of running coupling constants and the non-perturbative $\beta$ function for each coupling constant can be obtained by using FRG approach. Full effective action can be obtained as Eq.~(\ref{Eq:stru}) is solved.

\subsection{NUMERICAL IMPLEMENTATION}\label{Sec:frgnum}

In this article we take the Taylor expansion method to carry out the numerical calculations.
Solving Eq.~(\ref{Eq:stru}) requires initial conditions which are the parameters in the Lagrangian, i.e., Eq.~(\ref{Eq:lag}).
We make use of the parameters given in Ref.~\cite{Resch:2017} which are summarized in Table~\ref{Tab:ini}.
\begin{table}[htb]
\caption{The initial conditions for Eq.~(\ref{Eq:stru}).}
\begin{tabular}{cccccccc}
 \hline
\hline
 $g$  & $h_x/\Lambda^3$ & $h_y/\Lambda^3$ & $c_a/\Lambda$ & $a_{10}/\Lambda^2$ & $a_{20}$ & $a_{01}$ & $\Lambda/{\rm GeV}$ \\  \hline
 $6.5$  &~$0.121^3$~ & ~$0.336^3$~  &  ~$4.808$~ & ~$0.56^2$~ &  ~$26$~ & ~$50$~ & ~$1$ \\
\hline
\hline
\end{tabular}
\label{Tab:ini}
\end{table}

The first three parameters in Table~\ref{Tab:ini} are fixed by the Goldberger--Treiman relation and the PCAC theorem~\cite{Lenaghan:2000PRD}. $c_{a}^{}$ represents the strength of ${\rm U_{A}(1)}$ anomaly
and it depends generally on the external parameters such as temperature or magnetic field.
In this article we set $c_{a}^{}$ to a constant to simplify the discussions.
$\Lambda$ is the ultraviolet cutoff of PQM model and it's fixed by the result that gluon develops a mass gap
below $1\,{\rm GeV}$  and then decouples with low energy physics~\cite{Braun:2016PRD}.
Thus in practical calculations,  we tune only the parameters $a_{10}$, $a_{20}$ and $a_{01}$
to fit the hadron observables at vanishing temperature and magnetic field.
The fitted results of the hadron properties are listed in Table~\ref{Tab:obs}.
\begin{table}[htb]
\caption{Calculated hadronic observables (in MeV). $\kappa^\pm,\kappa^0$ and $\bar{\kappa}^0$ are degenerate as well as $K^\pm, K^0$ and $\bar K^0$.}
\begin{tabular}{cccccc}
  \hline
   \hline
~~$f_\pi$~~  & $f_K$ & $m_\pi$ & $m_K$ & $m_\sigma$ & $m_{f_0}$  \\  \hline
 ~~$92.0$~~  &~~$113.4$~~ & ~~$138.3$~~  &  ~~$495.1$~~ & ~~$601.8$~~ &  ~~$1234.7$~~  \\
\hline
\hline
~~$m_{\eta'}$~~  & $m_{\eta}$ & $m_{a_0}$ & $m_\kappa$ & $m_{u,d}$ & $m_s$  \\  \hline
 ~~$962.8$~~  &~~$539.1$~~ & ~~$1013.7$~~  &  ~~$1102.5$~~ & ~~$299.0$~~ &  ~~$438.1$~~  \\
\hline
\hline
\end{tabular}
\label{Tab:obs}
\end{table}

In the fitting process, the $f_{\pi}^{}$ and $f_{K}^{}$ are obtained via the minimums
of  $\Gamma_{k=0}$. Then the constituent masses of the light (u,d) quark and strange quark can be written as
\begin{eqnarray}
m_{u,d}=\frac{g}{2}f_\pi\ \ ,\ \ m_s=\frac{g}{2}(2f_K-f_\pi)\, .
\end{eqnarray}
The masses of the mesons are given by the eigenvalues of the Hessian matrix $H_{i,j}$
\begin{eqnarray}
H_{i,j}=\frac{\partial^2U_{k=0}}{\partial \sigma_i\partial \sigma_j}\, .
\end{eqnarray}
The specific form of  the Hessian matrix is quite complicated and can be found in Refs.~\cite{Kamikado:2015JHEP,Mitter:2014PRD}.

The Polyakov-loop potential $V_{poly}$ also needs to be specified for solving Eq.~(\ref{Eq:stru}). We use the polynomial form to parametrize $V_{poly}$~\cite{Ratti:2005PRD}, which reads
\begin{eqnarray}
\frac{V_{poly}(\Phi,\bar\Phi)}{T^4}=&&\ -\frac{b_2(T)}{2}\Phi\bar\Phi-\frac{b_3}{6}(\Phi^3+\bar\Phi^3)+\frac{b_4}{4}(\Phi\bar\Phi)^2 \, ,\nonumber\\
b_2(T)=&&\ a_1+\frac{a_2}{1+t}+\frac{a_3}{(1+t)^2}+\frac{a_4}{(1+t)^3}\, ,
\end{eqnarray}
where $t=(T-T_{\rm YM})/T_{{\rm YM}}$.  $T_{{\rm YM}}$ is the critical temperature for the deconfinement phase transition in pure Yang-Mills theory.  When quarks are added, $t_{\rm YM}$ needs to be rescaled to account for the unquenched effect
\begin{eqnarray}
t_{{\rm YM}}=\frac{T-T_{\rm YM}}{T_{\rm YM}} \Longrightarrow
\alpha\ t_{{\rm QCD}}=\alpha\frac{T-T_{\rm QCD}}{T_{\rm QCD}}\, ,
\end{eqnarray}
where $\alpha$ and $T_{\rm QCD}$ are obtained by fitting lattice QCD data.
In this work we take the parameters recommended in Ref.~\cite{Fu:20182} which are collected
in Table~\ref{Tab:poly}.
Note that we are working at zero chemical potential, the $\Phi$ remains thus real.
We set then $\bar\Phi=\Phi$ in the rest of this article.
\begin{table}[htb]
\caption{Parameters for Polyakov-loop potential $V_{poly}$.}
\begin{tabular}{cccccccc}
  \hline
   \hline
 $a_1$  & $a_2$ & $a_3$ & $a_4$ & $b_3$ & $b_4$ & $\alpha$ & $T_{\rm QCD}$ \\  \hline
 $6.75$  &~$-1.95$~ & ~$2.625$~  &  ~$-7.44$~ & ~$0.75$~ &  ~$7.5$~ & ~$0.54$~ & ~$250 \,{\rm MeV}$ \\
\hline
\hline
\end{tabular}
\label{Tab:poly}
\end{table}

There is one more thing we should mention. When the external parameters (such as $2\pi T$ or $eB$) are comparable to the cutoff $\Lambda\sim1\,{\rm GeV}$, the initial effective action $\Gamma_{k=\Lambda}$ will grow a dependence on these external parameters.
This dependence is omitted in the above discussion and we employ the method proposed in Refs.~\cite{Fu:2018,Fu:20182} to compensate for this neglected dependence.
More detailed discussions can be found in Ref.~\cite{Braun:2018}.
The basic idea is to start the flow from a high enough scale in which all the external parameter dependence
can be neglected, and then flow down to the original cutoff $\Lambda$. As a result, the initial condition $\Gamma_\Lambda$ will get modified
\begin{eqnarray}
\Gamma_\Lambda[0]\rightarrow\Gamma_\Lambda[0]+\Delta\Gamma_\Lambda[s]\, ,
\end{eqnarray}
where $s$ strands for all the external parameters ($T,eB$ in this article). $\Delta\Gamma_\Lambda[s]$ represents the external parameter effect and reads
\begin{eqnarray}\label{Eq:extmod}
\Delta\Gamma_{\Lambda}[s]\ &&=[\Gamma_{\Lambda}(s) - \Gamma_{\Lambda}(0)]  \nonumber\\
&&=[\Gamma_{\Lambda}(s) - \Gamma_{\infty}(s)] - [\Gamma_{\Lambda}(0) - \Gamma_{\infty} (0)] \nonumber\\
&&=\int^{\Lambda}_{\infty} dk\ [\partial_{k} \Gamma_{k}(s) - \partial_{k}\Gamma_{k}(0)] \, .
\end{eqnarray}
The second line follows from that external parameter dependence can be neglected when the ultraviolet cutoff is large enough, i.e., $\Gamma_{\infty}^{} (s) = \Gamma_{\infty} (0)$.
Note that mesons will become too heavy above $\Lambda$ and decouple from the system,
we only substitute the fermion part of Eq.~(\ref{Eq:effpotflow}) into Eq.~(\ref{Eq:extmod}).
For numerical convenience, we set the temperature to zero in the $\partial_{k} \Gamma_{k}(0)$ term of Eq.~(\ref{Eq:extmod}) and keep the magnetic field dependence. The specific form of Eq.~(\ref{Eq:extmod}) now reads
\begin{eqnarray}
\Delta U_{\Lambda}^{}[T,eB] = \int_{\infty}^{\Lambda} dk\ \frac{k^{4}}{12\pi^{2}}\sum_{f={u,d,s}}
\alpha_{f}(k)\ \frac{2\tilde{n}_{f}^{}(E_{f},\Phi)}{E_{f}}\, ,\nonumber\\
\end{eqnarray}
where $E_{f}=\sqrt{k^{2} + m_{f}^{2}}$.
Generally, quark mass $m_{f}$ depends on $\rho_{1}^{}$ and $\rho_{2}^{}$.
However it is shown in Ref.~\cite{Fu:2018} that setting $m_{f}$ to its constituent mass value is a good approximation. Thus we set $m_{u,d}=300\, {\rm MeV}$ and $m_{s}=430\, {\rm MeV}$ in our calculations.

\section{Calculations and Results}\label{Sec:res}

Once the effective action $\Gamma$ is known, many thermodynamic quantities of the system can be obtained.  Some quantities are easy to get while some others contain cutoff artifacts and should be handled carefully. We will describe how to obtain these quantities below.

First of all, the minimums of the $\Gamma(\rho_{1}^{},\rho_{2}^{},\Phi)$ are related to the order parameters
and we use $\tilde{\rho}_{1}^{},\tilde{\rho}_{2}^{},\tilde{\Phi}$ to denote them.
%
\iffalse $\tilde{\rho}_1$ and $\tilde{\rho}_2$ are related to the expectation values of $\tilde{\sigma}_x$ and $\tilde{\sigma}_y$ via
\begin{eqnarray}
\begin{split}
\tilde{\rho}_{1}^{}=&\ \frac{1}{2}(\tilde{\sigma}_{x}^{2} + \tilde{\sigma}_{y}^{2}) \, , \\
\tilde{\rho}_{2}^{}=&\ \frac{1}{24}(2\tilde{\sigma}_{y}^{2} - \tilde{\sigma}_{x}^{2})^{2} \, .
\end{split}
\end{eqnarray}
\fi
%
We take $\tilde{\rho}_{1}^{}$ as the order parameter for chiral phase transition because
$\tilde{\rho}_{1}^{}$ is related to the condensates of  light quark and strange quark, which reads
\begin{eqnarray}
\tilde{\rho}_{1}^{}=\frac{1}{2}(\tilde{\sigma}_x^2+\tilde{\sigma}_y^2)\, .
\end{eqnarray}
And it is usual to use $\tilde{\Phi}$ as the order parameter for deconfinement phase transition.
We define the pseudo-critical temperature as the inflection point of these order parameters where they change most rapidly.

The pressure $P$ and the entropy density $S$ can be derived directly from $\Gamma$ as
\begin{eqnarray}\label{Eq:therm}
P(T,eB)=&&\ -\frac{T}{V}\ \Gamma(\tilde{\rho}_1,\tilde{\rho}_2,\tilde{\Phi})\big|_{T,eB} \, , \nonumber\\
S(T,eB)=&&\ \frac{\partial P}{\partial T} \, .
\end{eqnarray}
Meanwhile, we can expand the pressure $P(T,eB)$ in terms of $eB$ as
\begin{eqnarray}\label{Eq:pressure}
P(T,eB)=P(T,0)+\frac{1}{2}\tilde{\chi}(T)(eB)^2+\mathcal{O}((eB)^4)\, , \qquad
\end{eqnarray}
where $\tilde{\chi}(T)$ is the bare magnetic susceptibility which is usually taken to characterize the leading response $\mathcal{O}((eB)^2)$ of the pressure to the magnetic field.
Unfortunately, both $P(T,0)$ and bare magnetic susceptibility $\tilde{\chi}(T)$ contain cutoff artifacts
and need to be renormalized.
We follow the procedures taken in Refs.~\cite{Bali:2014JHEP,Bonati:2014PRD,Levkova:2014PRL} to define the renormalized magnetic susceptibility as
\begin{eqnarray}
\chi(T) \equiv \tilde{\chi}(T) - \tilde{\chi}(0) \, ,
\end{eqnarray}
and $\chi(T)$ is obtained by a polynomial fitting to $\Delta P$~\cite{Kamikado:2015JHEP} in the weak magnetic field region ($eB<0.1\,{\rm GeV}^2$), which reads
\begin{eqnarray}   \label{eqn:MFinducedPressure}
\Delta P = &&\ [P(T,eB)-P(T,0)] - [P(0,eB) - P(0,0)]  \nonumber\\
=&&\ \frac{1}{2}\chi(T)(eB)^{2} + \mathcal{O}((eB)^{4}) \, .
\end{eqnarray}
Cutoff artifacts are all removed by the subtractions in $\Delta P$.

Magnetic susceptibility $\chi(T)$ has also been studied by combining the FRG approach with the non-interacting gas approximation in Refs.~\cite{Agasian:2008Review,Kamikado:2015JHEP}, in which the whole system is approximated as non-interacting quark and (pseudo-)scalar meson gases \cite{Endrodi:2013JHEP}. $\chi(T)$ can then be separated into two parts,
%
%and we extend the formulas therein to PQM model, it reads
%
namely
\begin{eqnarray}
\chi(T) = \chi_{q}^{}(T) + \chi_{m}^{}(T) \, ,
\end{eqnarray}
where $\chi_q(T)$ and $\chi_m(T)$ refer to the contributions from quarks and mesons respectively.
After some calculations, we have the two terms explicitly as
\begin{eqnarray}\label{Eq:sushrg}
\chi_{q}^{}(T) = \frac{N_c}{6\pi^2} \sum_{f} q_{f}^{2} \Bigg\{&&2\int_{0}^{\infty} dx\ \frac{\tilde{n}_{f}^{} (\sqrt{x^{2} +m_{f}^{2}(T)},\Phi(T))}{\sqrt{x^{2} + m_{f}^{2}(T)}} \nonumber\\
&& -{\rm ln}\bigg(\frac{m_{f}(0)}{m_{f}(T)}\bigg)\Bigg\}\, ,\nonumber\\
\chi_{m}(T) = - \frac{1}{48\pi^2}\sum_{b} \Bigg\{&&2\int_{0}^{\infty} dx\ \frac{n_{b}^{} (\sqrt{x^{2} + m_{b}^{2}(T)})}{\sqrt{x^{2} + m_{b}^{2}(T)}}\nonumber\\
&& +{\rm ln}\bigg(\frac{m_{b}^{}(0)}{m_{b}^{}(T)}\bigg)\Bigg\}\, ,
\end{eqnarray}
where the summation for bosons is over charged mesons ($\pi^{+}$, $\pi^{-}$, $K^{+}$, $K^{-}$, $a_{0}^{+}$, $a_{0}^{-}$, $\kappa^{+}$, $\kappa^{-}$). Note that the masses of the particles have a dependence on the temperature and neglecting this dependence will lead to a negative definite magnetic susceptibility at any temperature.
%
%Ref.~\cite{Kamikado:2015JHEP} proposes to use the temperature-dependent masses and then includes additional logarithm terms in Eq.~(\ref{Eq:sushrg}) which are vanished
%
It is proposed in Ref.~\cite{Kamikado:2015JHEP} to use the temperature-dependent masses and then include additional logarithm terms in Eq.~(\ref{Eq:sushrg}), which are vanishing
in the original calculation. These temperature-dependent masses $m_{f(b)}(T)$ and Polyakov-loop $\Phi(T)$ can be obtained from FRG approach, we can thus substitute the results into the above formulas and do the integrals numerically.

To check our calculations further, Adler function \cite{Adler:1974PRD,Baikov:2010PRL}  is also calculated. It's mentioned in Ref.~\cite{Bali:2014JHEP} that $\chi(T)$ is closely related to the Adler function which can be calculated in perturbation theory.  The relation can be formulated as
\begin{eqnarray}
D(2\pi T)=6\pi^2T\frac{\partial \chi(T)}{\partial T}\, ,
\end{eqnarray}
where $D(2\pi T)$ is the Adler function and the momentum $Q$ of virtual photon is replaced by Matsubara frequency $2\pi T$. This relation enables us to compare our results with those given in perturbation theory.

 In the following subsections, we will represent and discuss our main results.
 %
\iffalse
\begin{eqnarray}
\tilde{\chi}(T)=&&\ \frac{\partial^2 P}{\partial (eB)^2}=\frac{1}{\beta Ve^2}\int d^4xd^4y \ \frac{\delta^2 W[A]}{\delta A_\mu(x)\delta A_\mu(y)}\nonumber\\=&&\ \frac{1}{e^2}\int d^4x\ \frac{\delta^2 W[A]}{\delta A_\mu(x)\delta A_\mu(0)}\nonumber\\=&&\sum_f\ q_f^2\int d^4x\ \langle 0|T[j^f_\mu(x)j^f_\mu(0)]|0\rangle,
\end{eqnarray}
where $j_{\mu}^{f}$ denotes the current operator for f quark and $q_f$ is the corresponding charge number. And the current-current correlation function in the last line is related to Adler function in the usual way.
\fi

\subsection{ORDER PARAMETERS AND PHASE DIAGRAM}

\begin{figure}[htb]
\centering
\includegraphics[width=0.43\textwidth]{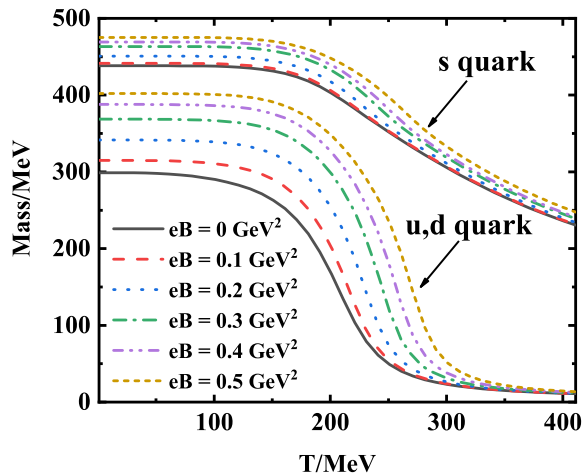}
\vspace*{-3mm}
\caption{(color online) Calculated constituent masses of light quark and strange quark as functions of temperature at several values of magnetic field strength.}\label{fuuds}
\end{figure}

\begin{figure}[htb]
\centering
\includegraphics[width=0.43\textwidth]{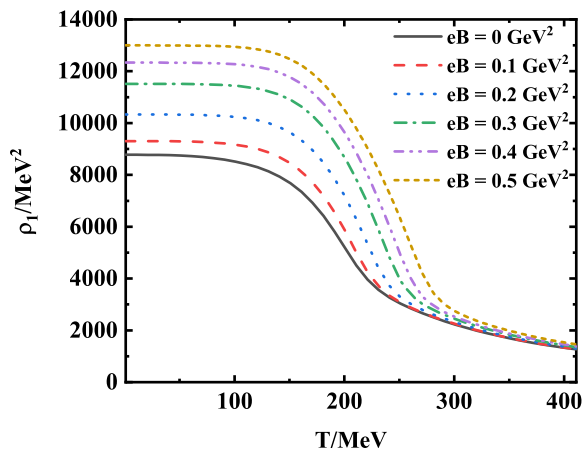}
\vspace*{-3mm}
\caption{(color online) Calculated chiral order parameter $\rho_1$ at several values of magnetic field strength.}\label{furho1}
\end{figure}

The calculated results of the constituent quark masses as functions of temperature at several values of magnetic field strength for the light quark and strange quark are illustrated in Fig.~\ref{fuuds}.
We show that, at a fixed temperature of any value,
the constituent quark mass increases with the ascending of the magnetic field strength.
This phenomenon is just the so called magnetic catalysis and is explained by the dimension reduction of quarks under external magnetic field, see, e.g., Ref.~\cite{Miransky:2002PRD}.
In order to take the condensate of  light quark and strange quark  both into consideration,
we use the $\rho_{1}^{}$ as the order parameter for chiral phase transition and
the obtained variation behavior against the temperature is illustrated
in Fig.~\ref{furho1}.
The $T_c^\chi$ extracted from the inflection point of the $\rho_{1}^{}$ at $eB=0$ is $T_{c}^{\chi}=199\, {\rm MeV}$.
Figs.~\ref{fuuds} and \ref{furho1} both show evidently that the 2+1 flavor PQM model with FRG approach
still displays the character of magnetic catalysis and more sophisticated models are needed to realize
the inverse magnetic catalysis.
\begin{figure}[htb]
\centering
\includegraphics[width=0.43\textwidth]{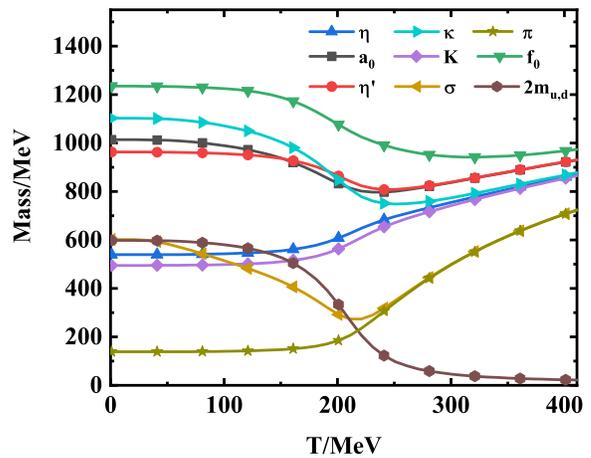}
\vspace*{-3mm}
\caption{(color online) Calculated meson masses and twice light quark mass as functions of temperature at $eB=0$. $\kappa^\pm,\kappa^0$ and $\bar{\kappa}^0$ are degenerate as well as $K^\pm, K^0$ and $\bar K^0$}\label{fumass}
\end{figure}
\begin{figure}[htb]
\centering
\includegraphics[width=0.45\textwidth]{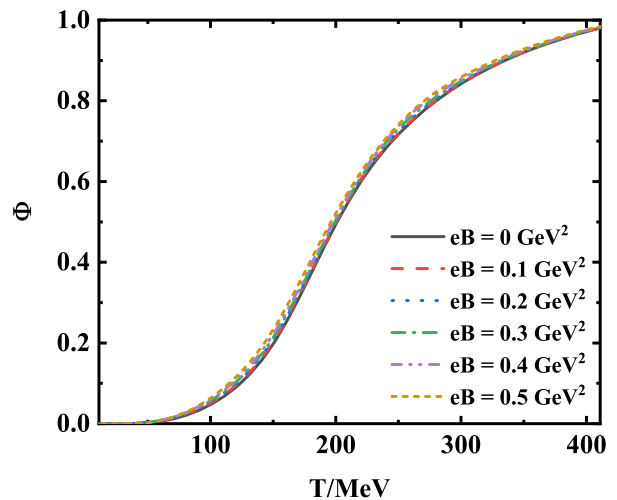}
\vspace*{-3mm}
\caption{(color online) Calculated Polyakov-loops as functions of temperature at several values of magnetic field strength.}\label{fupoly}
\end{figure}
\begin{figure}[htb]
\centering
\includegraphics[width=0.40\textwidth]{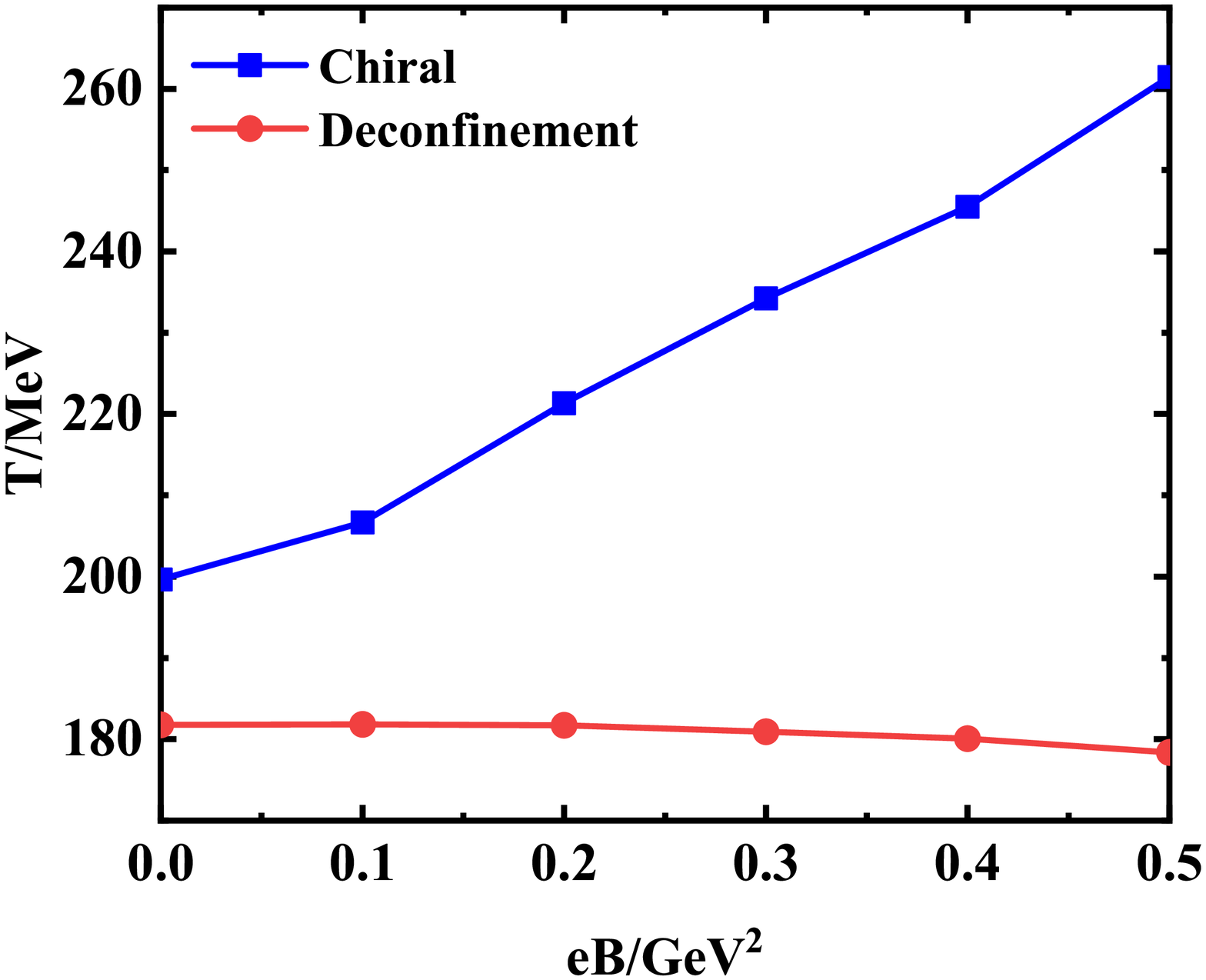}
\vspace*{-3mm}
\caption{(color online) Calculated QCD phase diagram for the chiral phase transition (line with squares)
and the deconfinement phase transition (line with circles) in $T$--$eB$ plane.}
\label{fuphase}
\end{figure}

The restoration of chiral symmetry can also be seen from the temperature dependence of the mass spectrum of mesons displayed in Fig.~\ref{fumass}.
We can see evidently that the masses of $\pi$ and $\sigma$ mesons get close as the temperature increases and become finally degenerate, which signals the restoration of chiral symmetry.
It is also apparent that the mesons become much heavier than two quarks when chiral symmetry is restored,
this means that the mesons become unphysical degrees of freedom as the chiral symmetry is restored
since the mesons have in fact melted to quarks and antiquarks.

As usual, we take the Polyakov-loop $\Phi$ as the order parameter for deconfinement phase transition and the  corresponding pseudo-critical temperature $T_{c}^{d}$ is defined by the inflection point of the $\Phi$.
The obtained temperature dependence of the $\Phi$ at several values of magnetic field strength is shown in Fig.~\ref{fupoly}.
The obtained $T_{c}^{d}$ in case of vanishing magnetic field is $182\,$MeV.
And the $T_{c}^{d}$ decreases slightly as the magnetic field strength increases,
which is consistent with the result given in Ref.~\cite{Andersen:2014JHEP}.
Such a slightly decreasing behavior of the $T_{c}^{d}$ can be explained by the energy change of quarks.
When the magnetic field is turned on, the dispersion relation of the quarks get modified
\begin{eqnarray}\label{Eq:disp}
E^{2} = m^{2} + \vec{\,p}^2 \Longrightarrow m^{2} + p_{z}^{2} + (2l+1-s) eB \, ,
\end{eqnarray}
where $l$ labels the Landau level and $s=\pm1$ denotes the spin orientation of the particle.
Note that quark can stay at the lowest Landau level ($l=0$ and $s=1$) and its energy is reduced,
that's exactly how dimension reduction happens.
It will take less energy to create a quark from vacuum, then the Polyakov-loop $\Phi$ increases since Polyakov-loop $\Phi\sim e^{-\beta F_q}$ where $F_q$ is the free energy of a single-quark system~\cite{McLerran:1981PRD}.
And in turn, the $T_c^d$ gets lowered.
We will discuss the response of different particles to the magnetic field further in the following subsections.  Here, we would like to mention that, since the magnetic field has not yet been included explicitly in the Polyakov-loop potential $V_{poly}$ in this work, the external magnetic field affects the Polyakov-loop rather slightly, which is consistent with what has been given in Ref.~\cite{Andersen:2014JHEP}.

In Fig.~\ref{fuphase}, we display the obtained magnetic field strength dependence of the $T_{c}^{d}$ and $T_{c}^{\chi}$. It indicates clearly that the $T_{c}^{\chi}$ and $T_{c}^{d}$ split from each other
as the magnetic field increases.
Considering the increasing amplitude of the $T_{c}^{\chi}$ with respect to the magnetic field strength,
our present result is almost exactly the same as the previous result in FRG approach~\cite{Kamikado:2014JHEP}
in which $T_{c}^{\chi}$ increases about $40\,{\rm MeV}$ at
$eB = 20\, m_{\pi}^{2}$.
And in mean-field approximation
(e.g., in QM model), the increasing amplitude is a little mild,
Ref.~\cite{Andersen:2016Review} gives about $30\, {\rm MeV}$ increasement at $eB = 20\, m_\pi^2$.

\subsection{PRESSURE AND ENTROPY DENSITY}

\begin{figure}[htp]
\centering
\includegraphics[width=0.43\textwidth]{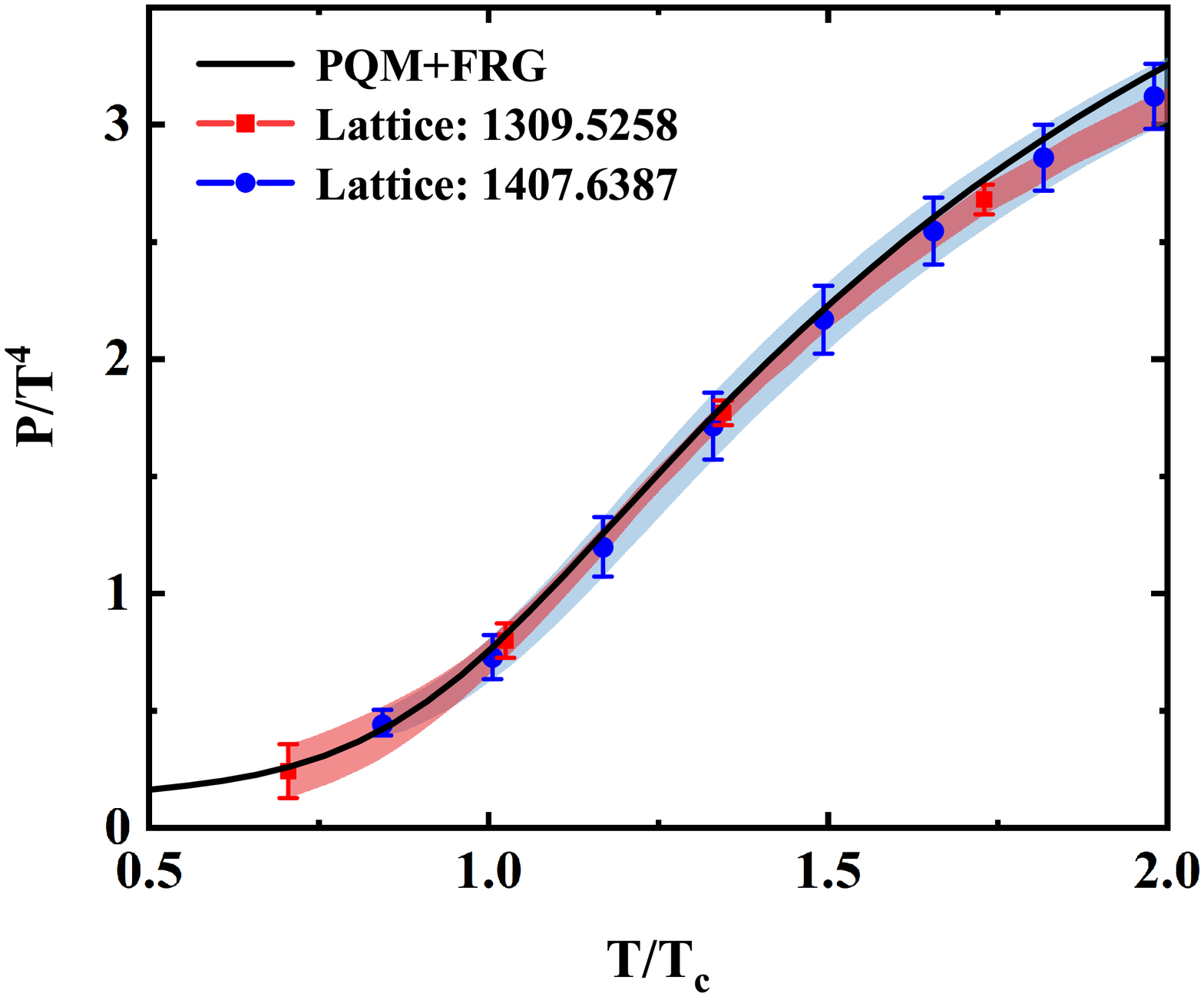}
\includegraphics[width=0.43\textwidth]{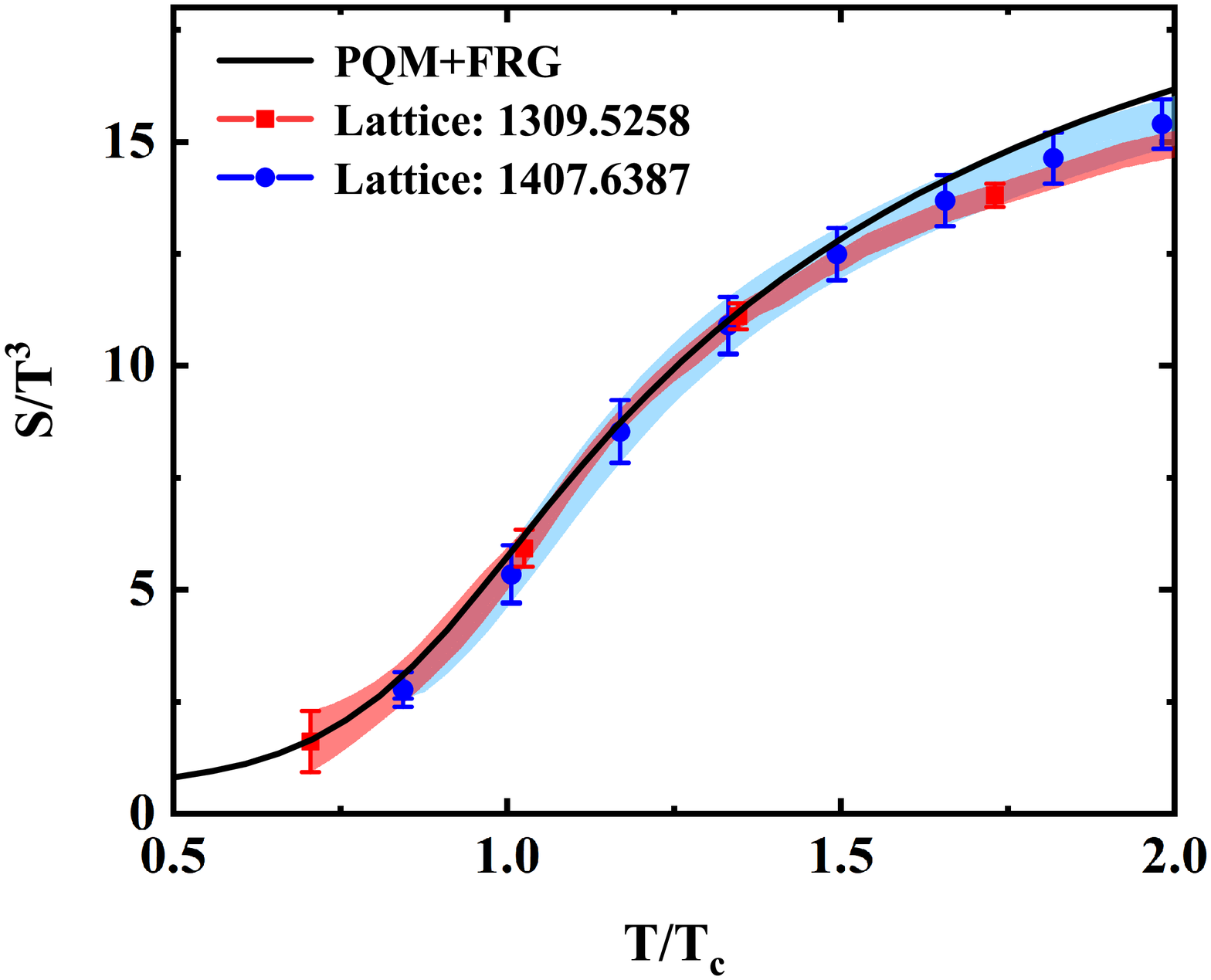}
\vspace*{-3mm}
\caption{(color online) Calculated normalized pressure (\emph{upper panel}) and normalized entropy density (\emph{lower panel}) as functions of normalized temperature at $eB=0$. Red and blue bands are the results from lattice QCD  simulations~\cite{Bazavov:2012PRD,Borsanyi:2010JHEP}.}
\label{futher}
\end{figure}

In Fig.~\ref{futher}, we show the calculated normalized pressure and normalized entropy density at $eB=0$
and compare them with the lattice QCD simulation results~\cite{Bazavov:2012PRD,Borsanyi:2010JHEP}.
In order to make a model-independent comparison, the temperature axis is normalized by the $T_c^\chi$ calculated in lattice QCD ($T_{c}^{\chi}=154\, {\rm MeV}$~\cite{Bazavov:2012PRD}) and the present PQM + FRG approach ($T_{c}^{\chi}=199\, {\rm MeV}$), respectively.
We can see from the figure that our results are in reasonable agreement with lattice QCD results
after the temperature normalization is made.
The result shows evidently that the entropy density increases generally with the rising of temperature.
Furthermore, the monotonicity of the normalized entropy density function in terms of
the temperature changes from up-concave to up-convex as the temperature increases to a specific value.
With the chiral pseudo-critical temperature fixed by analyzing the quark condensate (and the constituent quark mass) in mind,
one can notice that such a inflection temperature for the normalized entropy density is almost the same as the
$T_{c}^{\chi}$ for the chiral symmetry to be restored.
\begin{figure}[htb]
\centering
\includegraphics[width=0.45\textwidth]{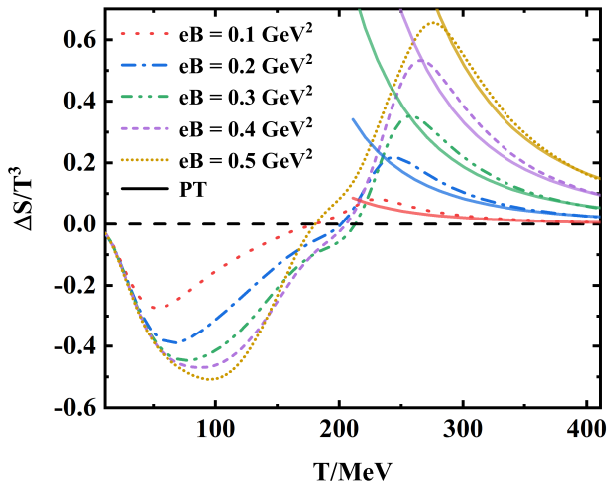}
\vspace*{-3mm}
\caption{(color online) Calculated normalized magnetic field induced entropy density $[S(T,eB)-S(T,0)]/T^3$ in terms of  temperature, solid lines in the figure are the Stefan-Boltzmann limits for each magnetic field. }
%And normalized entropy density at low temperature region (\emph{lower panel}).}
\label{fuentro}
\end{figure}

When the magnetic field is turned on, many thermodynamic quantities (such as pressure) will develop
a dependence on the $eB$  and consequently their normalized values will diverge at low temperature~\cite{Bali:2014JHEP}. However, entropy density $S$ doesn't contain such a divergence and $S$ need not to be renormalized.
 %
 %thanks to the derivative with respect to $T$ in Eq.~(\ref{Eq:therm}).
%
We will thus analyze the normalized magnetic field induced entropy density $[S(T,eB)-S(T,0)]/T^3$ in the following.

The calculated variation behaviors of the normalized magnetic field induced entropy density in terms of the  temperature at several values of the magnetic field strength are displayed in Fig.~\ref{fuentro}.
One can recognize easily from Fig.~\ref{fuentro} that the entropy density decreases with the increasing of the magnetic field strength in the temperature region in which the chiral symmetry is dynamically broken.
Nevertheless, it increases as the magnetic field strength gets stronger when the chiral symmetry is restored in high temperature region, which is qualitatively consistent with the results of lattice QCD~\cite{Bali:2014JHEP} and HRG model~\cite{Endrodi:2013JHEP}.
This behavior can be explained as below.
From Figs.~\ref{fuuds} and \ref{furho1}, we can see that the magnetic field strengthens the quark condensate drastically at low temperature.
It implies that the QCD system at low temperature selects out the microscopic states whose condensate
doesn't vanish when the magnetic field gets stronger. This selection makes the system more ordered and, in turn,
the entropy density decreases.
At high temperature, this selection becomes less important and other effects should be considered
to account for the increase of the entropy density.
More precisely speaking, if we approximate the system as non-interacting gases,
the entropy of the PQM system receives contributions from both quarks and mesons.
Combining Eqs.~(\ref{Eq:therm}) and (\ref{Eq:sushrg}), we can see that the entropy density of the charged (pseudo-)scalar mesons (such as $\pi^{\pm}\, , \cdots$) and free quarks response differently to the magnetic field: the energy of each $\pi^\pm$ meson is increased by the magnetic field (see Eq.~(\ref{Eq:disp})),
thus there are less microscopic states available to the $\pi^{\pm}$ system at certain temperature,
which leads to a lower entropy density;
while the energy of each quark is decreased, hence the entropy of quark system increases.
Since the induced entropy of the whole system is mainly determined by the lightest charged mesons $\pi^\pm$ when the chiral symmetry is broken, the entropy density decreases then at low temperature as the magnetic field gets stronger.
When the temperature is high enough, chiral symmetry is restored and mesons get melted.
The entropy density of the system is determined mainly by quarks
and thus increases with the increasing of the magnetic field.
We will try to discuss these behaviors from a more fundamental point of view in Sec.~\ref{Sec:sus}.

Moreover, we should note that the calculated entropy density at high temperature ($\gtrsim120\, {\rm MeV}$)
still shows quantitative difference with that given in lattice QCD simulations \cite{Bali:2014JHEP}.
Our presently calculated normalized magnetic field induced entropy density at different magnetic field strengths are quite small compared with the results given in Ref.~\cite{Bali:2014JHEP}.
This may be due to the missing of the inverse magnetic catalysis effect in PQM model.
More sophisticated model is required to solve this discrepancy.

\subsection{MAGNETIC SUSCEPTIBILITY AND ADLER FUNCTION}\label{Sec:sus}

\begin{figure}[htb]
\centering
\includegraphics[width=0.43\textwidth]{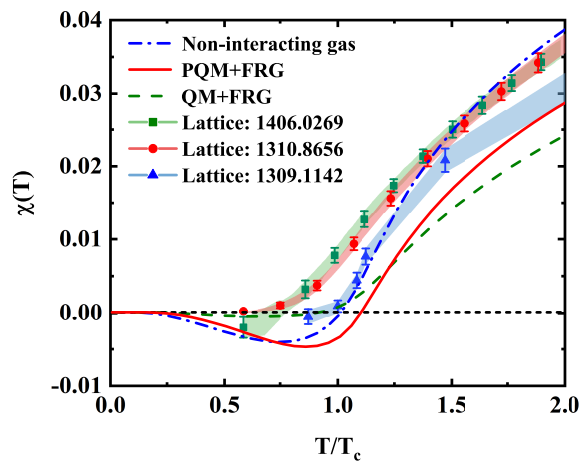}
\vspace*{-3mm}
\caption{(color online) Calculated magnetic susceptibility $\chi(T)$ from FRG approach (solid and dashed lines), non-interacting gas approximation (dot-dashed line) and lattice QCD simulations~\cite{Bali:2014JHEP,Bonati:2014PRD,Levkova:2014PRL} (colored bands). Different chiral pseudo-critical temperatures ($T_c^{\rm Lattice}=154\, {\rm MeV}$, $T_c^{\rm QM}=171\, {\rm MeV}$ and $T_c^{\rm PQM}=T_c^{\rm Gas}=199\, {\rm MeV}$) are used respectively in rescaling the temperature.}\label{fusus}
\end{figure}

We collect the $\chi$s obtained through the present (P)QM+FRG approach,
non-interacting gas approximation (see Eq.~(\ref{Eq:sushrg}))
and lattice QCD simulations together in Fig.~\ref{fusus}. The QM+FRG approach here also takes the parameters in Table~\ref{Tab:ini} except that the Polyakov-loop and its potential are removed.
The non-interacting gas approach approximates the whole system as non-interacting quark and (pseudo-)scalar meson gas, and the particles contribute individually to the magnetic susceptibility~\cite{Kamikado:2015JHEP} as shown in Eq.~(\ref{Eq:sushrg}).
Moreover, the temperature-dependent particle masses and Polyakov-loop in Eq.~(\ref{Eq:sushrg}) are taken from
the present PQM+FRG results (as displayed in Figs.~\ref{fumass} and \ref{fupoly}). 
The figure manifests apparently that, at temperature below $T_c^\chi$, $\chi$ is negative and QCD system is diamagnetic. This is because $\chi$ receives contributions mainly from the lightest pseudo-scalar mesons $\pi^\pm$ below $T_c^\chi$. It's well known that spin-less charged boson displays diamagnetic character due to its orbital motion perpendicular to the magnetic field.
When the temperature exceeds $T_{c}^{\chi}$, mesons become too heavy and decouple from the system,
in fact, get melted to quarks. $\chi$ will be solely determined by quarks then. External magnetic field will align quark's magnetic moment parallel to it and therefore leads to paramagnetic at high temperature. We can see from the above discussions that the transition of the magnetism of QCD system is closely related to the chiral symmetry, therefore the transition temperature of magnetism is near $T_{c}^{\chi}$.
Considering Eq.~(\ref{eqn:MFinducedPressure}) and the consistency between our present result and those given in lattice QCD simulations~\cite{Bali:2014JHEP,Bonati:2014PRD,Levkova:2014PRL}, one can recognize that our presently obtained magnetic field induced pressure also agrees with the lattice QCD result~\cite{Levkova:2014PRL} very well.

Comparing the results via the FRG approach embodied with the PQM and the QM models,
we can see that PQM model predicts a higher $\chi$ than QM model in the temperature region
above the $T_{c}^{\chi}$ and a much lower $\chi$ below the $T_{c}^{\chi}$.
This is because of the effect of Polyakov-loop.
Quarks are confined at low temperature in PQM model and their contributions to $\chi$ are suppressed,
which leads to a lower $\chi$.
At high temperature, gluon's effects are taken into consideration in PQM model through the Polyakov potential $V_{poly}$, which can account for the higher $\chi$.
The difference between the PQM and the QM models indicates that  the effect of Polyakov-loop is important
for calculating the magnetic susceptibility quantitatively.
We also note that the result from non-interacting gas approximation seems
agree with lattice QCD simulations quite well,
but the increasing rate is larger than those given in lattice QCD at the temperature higher than the $T_{c}^{\chi}$.
%
%non-interacting gas approximation with temperature-dependent masses works well in this case.}
%

\begin{figure}[htb]
\centering
\includegraphics[width=0.43\textwidth]{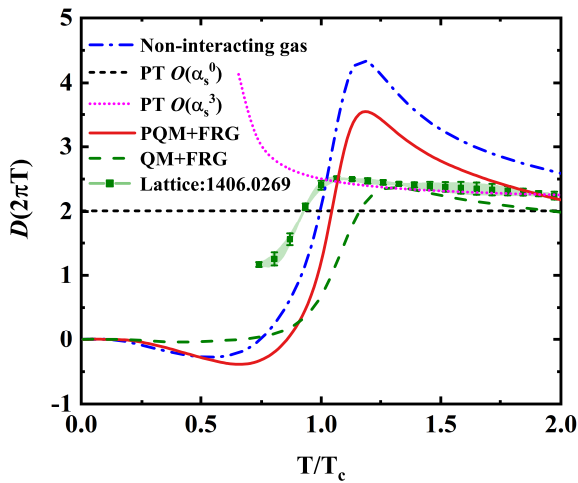}
\vspace*{-3mm}
\caption{(color online) Calculated Adler function $D(2\pi T)$ from FRG approach (solid and dashed lines),
non-interacting gas approximation (dot-dashed line), lattice QCD simulations~\cite{Bali:2014JHEP} (colored band) and perturbation theory~\cite{Baikov:2010PRL} (short dashed and dotted lines).
Temperatures are rescaled by the corresponding chiral pseudo-critical temperatures:
$T_{c}^{\rm QM}=171\,{\rm MeV}$, $T_{c}^{\rm Lattice}=T_{c}^{\rm PT}=154\,{\rm MeV}$
and $T_{c}^{\rm PQM}=T_{c}^{\rm Gas}=199\, {\rm MeV}$. }
\label{fuadler}
\end{figure}

To check our calculation results and compare with the lattice QCD simulation result further,
we show the calculated Adler function
and the comparison with the perturbative QCD result~\cite{Baikov:2010PRL} and lattice QCD result \cite{Bali:2014JHEP} in Fig.~\ref{fuadler}.
The figure shows clearly that, besides the reasonable agreement between our presently calculated result and the perturbative QCD result in high temperature region, the drastic variation behavior around the pseudo-critical temperature also coincides with the lattice QCD result. We also note that the calculated Adler functions from PQM and QM model behave differently.
QM model only consists of free quarks at high temperature, thus its Adler function is close to the perturbative result at the lowest order $\mathcal{O}(\alpha_{s}^{0})$.
While quarks in PQM model are still coupled with gluons at high temperature by the effect of Polyakov-loop,
thus the corresponding Adler function is more close to the higher order result at high temperature.

Before we close this section, we try to explain the response of QCD system to the external magnetic field in an intuitive picture. At low temperature, chiral symmetry is dynamically broken and quarks prefer to come into pairs, then the whole system is composed of various quark pairs (in fact, quark--antiquark pairs, here).
When the magnetic field is turned on, these quark pairs will rotate in the plane perpendicular to
the magnetic field in a way that produces anti-parallel magnetic moment to the external field,
which leads to a negative magnetic susceptibility.
Interestingly, these behaviors of quarks have many similarities with the electrons in BCS theory~\cite{Bardeen:1957PR}, in which electrons form Cooper pairs and lead to Meissner effect.
Moreover, these quark pairs can be equivalently regarded as various mesons and we can see from the previous discussions that the entropy of these quark pairs decreases with the increasing of the magnetic field.
At high temperature, the chiral symmetry gets restored.
Quarks stop forming pairs and the whole system contains only quarks.
Then, external magnetic field will align the magnetic moment of the quarks parallel to it
and leads to a positive magnetic susceptibility.
The increasing feature of the the positive magnetic susceptibility with respect to the temperature means
that there exist more magnetic moments need to be aligned.
As a consequence, the entropy density of the quark system increases
with the increasing of the magnetic field strength.
These characteristics can be regarded as an intuitive picture for the magnetic catalysis
in simple phenomenological models.

\section{Summary and Remarks}\label{Sec:sum}

In this paper, we investigated the thermodynamical properties of the strong interaction matter in an external magnetic field by applying the FRG approach to (2+1) flavor PQM model.
Our calculation results of the temperature dependence of the pressure and the entropy density
are qualitatively consistent with the lattice QCD result while quantitative discrepancy still exists when the magnetic field is turned on. The calculated constituent masses of light quark and strange quark indicate that the pseudo-critical temperature for the chiral symmetry to be restored increases with the increasing of the magnetic field strength, which is consistent with the magnetic catalysis effect.

We have also studied the response of the QCD system to the magnetic field
by analyzing the magnetic field induced entropy density, the magnetic susceptibility and Adler function.
Our calculated results indicate that the property of QCD system undergoes a transition with the increasing of temperature: at temperature below $T_{c}^{\chi}$, the magnetic field reduces the entropy density and
the pressure of the system;
after chiral symmetry is restored,  the magnetic field begins to increase the entropy density
and the pressure.
This transition is closely related to the chiral symmetry breaking and its restoration.
Especially, an intuitive picture for the magnetic catalysis in simple phenomenological models is given.

In addition, because the model we have taken is simple so that the gauge field effects and the feedback effects
of the magnetic field on the gauge interaction (see, e.g., Ref.~\cite{Farias:2014PRC}) have not yet been taken into account,
our present result is then valid only to the pure and direct effect of the magnetic field on the strong interaction matter, but could not rule out the possibility of the inverse magnetic catalysis effect, at least the retarded magnetic catalysis~\cite{Ilgenfritz:2014PRD,Braun:2016PLB,Pawlowski:2015PRD}.
In other words, our investigation indicates that the underlying physics responsible for
the inverse magnetic catalysis (or magnetic inhibition) is beyond the physics demonstrated in the PQM model and other similar phenomenological models.
To include the gauge field effects and explore the physics for the magnetic inhibition,
one may take the nonlocal NJL model \cite{Pagura:2017PRD}, or the Dyson--Schwinger equations of QCD~\cite{Pawlowski:2015PRD},
or the approaches taking the feedback effect of the magnetic field on the strong interaction to carry out the calculations.
The related investigation is under progress.

\begin{acknowledgments}
The work was supported by the National Natural Science Foundation of China under Contracts No.\ 11435001,
No. 11775041, and the National Key Basic Research Program of China under Contract No.\ 2015CB856900.
\end{acknowledgments}

\end{CJK*}

\end{document}